%% file: root.tex
\documentclass[letter, 10pt, conference, hidelinks]{ieeeconf}      

\IEEEoverridecommandlockouts                              

\usepackage{graphicx} 
\usepackage{amsmath} 
\usepackage{amssymb}  
\usepackage{hyperref}
\usepackage{soul}
\usepackage{float}

  \usepackage{amsthm}
\theoremstyle{definition}
\newtheorem{defi}{Definition}
\theoremstyle{definition}
\newtheorem{asu}{Assumption}
\theoremstyle{plain}
\newtheorem{theorem}{Theorem}
\theoremstyle{definition}
\theoremstyle{definition}

\theoremstyle{definition}
\newtheorem{prob}{Problem}

\usepackage{xcolor}
\usepackage{easyReview}

\title{\LARGE \bf
Leader-Follower Formation Tracking Control of Quadrotor UAVs\\Using Bearing Measurements 
}

\author{Sander Doodeman$^{1}$, Zhiqi Tang$^{2}$, Marcelo Jacinto$^{3}$, Rita Cunha$^{3}$, and Carlos Silvestre$^{3,4}$
\thanks{$^{1}$Department of Mechanical Engineering, TU/e, Eindhoven University of Technology, Eindhoven, The Netherlands. E-mail:
        {\tt\small s.doodeman@tue.nl}}%
\thanks{$^{2}$Division of Decision and Control Systems, KTH Royal Institute of Technology, Sweden. E-mail:
        {\tt\small ztang2@kth.se}} %
\thanks{$^{3}$Institute for Systems and Robotics (ISR), Laboratory of Robotics and Engineering Systems (LARSyS), Instituto Superior Técnico, University of Lisbon, Portugal. E-mails:
        {\tt\small \{mjacinto, rita\}@isr.tecnico.ulisboa.pt }} %
\thanks{$^{4}$Department of Electrical and Computer Engineering of the Faculty of Science and Technology of the University of Macau, China. E-mail:
        {\tt\small csilvestre@umac.mo}} %
\thanks{The work of M. Jacinto was supported by the PhD Grant 2022.09587.BD from Funda\c{c}{\~a}o para a Ci{\^e}ncia e a Tecnologia
	(FCT), Portugal.}
}

\begin{document}

\maketitle
\thispagestyle{empty}
\pagestyle{empty}

\input{Abstract.tex}

\section*{Supplementary Material}
\noindent\textbf{Simulation Video:} \href{https://youtu.be/zqNK-d2ZgY0}{\texttt{youtu.be/zqNK-d2ZgY0}} \\
\noindent\textbf{Experiments Video:} \href{https://youtu.be/-cPlcVHDzzU}{\texttt{youtu.be/-cPlcVHDzzU}} \\
\noindent\textbf{Code:} \href{https://github.com/SDoodeman/bpe_quadrotor}{\texttt{github.com/SDoodeman/bpe\_quadrotor}}

\input{Introduction.tex}

\input{Preliminaries.tex}

\input{Modeling.tex}

\input{HierarchicalStrategy.tex}

\input{BPEformation.tex}

\input{Simulations.tex}

\input{Experiments.tex}

\input{Conclusion.tex}

 

\bibliographystyle{IEEEtran}
\bibliography{IEEEabrv,references}

\end{document}

%% file: Abstract.tex
\begin{abstract}
This work addresses the practical problem of distributed formation tracking control of a group of quadrotor vehicles in a relaxed sensing graph topology with a very limited sensor set, where only one leader vehicle can access the global position. Other vehicles in the formation are assumed to only have access to inter-agent bearing (direction) measurements and relative velocities with respect to their neighbor agents. A hierarchical control architecture is adopted for each quadrotor, combining a high-gain attitude inner-loop and an outer-loop bearing-based formation controller with collision avoidance augmentation. The proposed method enables a group of quadrotors to track arbitrary bearing persistently exciting desired formations, including time-varying shapes and rotational maneuvers, such that each quadrotor only requires relative measurements to at least one neighboring quadrotor. The effective performance of the control strategy is validated by numerical simulations in MATLAB and real-world experiments with three quadrotors.
\end{abstract}

%% file: Introduction.tex
\section{Introduction} \label{sec:intro}

The demand for a swarm of Unmanned Aerial Vehicles (UAVs) is rapidly expanding due to their ability to perform various missions, such as agricultural monitoring, exploration of large spaces, and search and rescue in disaster environments in a distributed and efficient manner \cite{chung2018survey}.
During these missions, it is essential for a swarm of UAVs to cooperatively track desired trajectories.
Traditional UAV formation tracking problems assume that the global position information is available for each agent, typically relying on accurate Global Navigation Satellite Systems (GNSS) to provide this information \cite{dong2016time}. While this assumption generally holds in open outdoor environments, GNSS signals are often unreliable in congested areas or unavailable altogether in indoor settings. In addition, in safety-critical missions, GNSS jamming and spoofing is a major concern. Therefore, resorting to onboard exteroceptive sensors is a much more effective solution to provide local relative measurements for UAVs. For instance, ultra-wideband delivers accurate relative distance measurements through radio communication, and vision sensors provide simple visual cues such as relative bearing (direction) measurements, which are robust to noise.
Due to the passive property of cameras, bearing-based formation control strategies can be established under sensing-only graph topologies, preferable when communication between vehicles is not available.

Due to the minimum sensing requirement,  bearing-based formation control has received growing attention.
Theoretical works \cite{zhao2016bearing,eren2003sensor} have exploited \textit{bearing rigidity} theory and formally described under which topological conditions the shape of a formation, up to a scaling factor, can be uniquely determined by constant inter-agent bearing measurements.
Based on the notion of classical \textit{bearing rigidity}, formation controllers relying on bearings have been designed for multi-agent systems under single-integrator dynamics \cite{zhao2016bearing, trinh2018bearing} and double-integrator dynamics \cite{zhao2019bearing} with applications to robotic vehicles.
For instance, a leader-follower formation tracking controller is proposed in \cite{zhao2019bearing} for a group of unicycle robots to track a translating rigid formation under the assumption that two leaders in the formation know their absolute global position. Many other works also require two leader vehicles \cite{parada2024twoleaders, ding_dynamics_2024, chen2023twoleaders, xu2020affine}.
The work in \cite{schiano2016rigidity} proposed a bearing-based formation stabilizing controller for quadrotors under simplified kinematic models, while \cite{erskine2021model} extended this approach to dynamic models using model predictive control. In both \cite{schiano2016rigidity} and \cite{erskine2021model}, a known distance measurement between two vehicles is necessary. Furthermore, just as in \cite{schiano2016rigidity}, most control strategies require classical bearing rigid conditions and are limited to achieving desired formations with constant bearing references only \cite{zhao2016bearing, trinh2019bearing}.

\begin{figure}
	\centering
        \vspace{0.2cm}
	\includegraphics[width=0.33\textwidth]{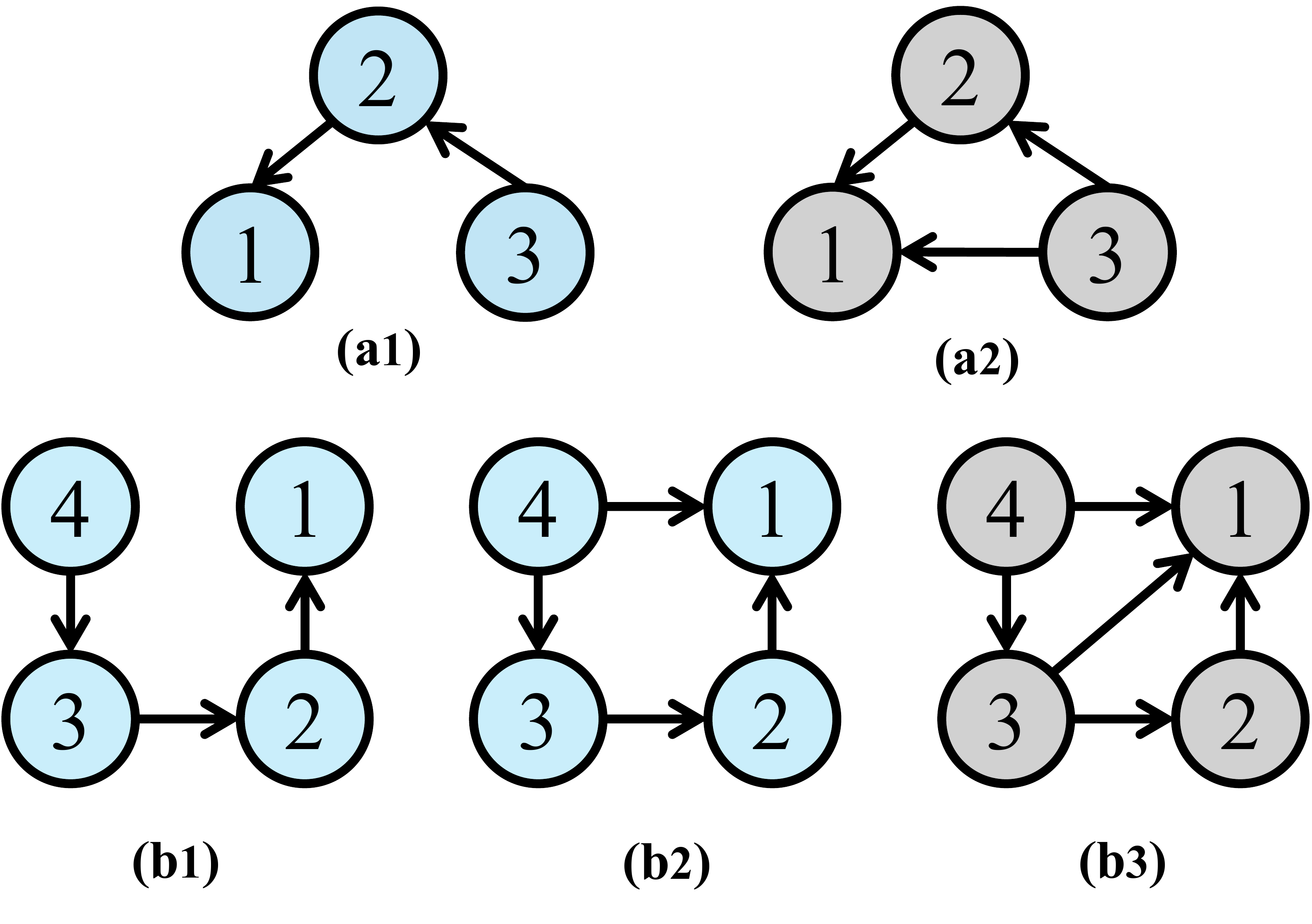}
	\caption{Possible graph topologies of leader follower BPE formation (a1)-(b3). Only (a2) and (b3) satisfy the minimum graph requirement by bearing rigid formation \cite{trinh2018bearing}.}
	\label{fig:BPE}
\end{figure}

Recently, Tang et al. \cite{tang2022relaxed, tang2021formation} introduced the concepts of \textit{bearing persistently exciting} (BPE) formation and \textit{relaxed bearing rigidity}, which provide bearing formation control solutions that can deal with time-varying bearing references while also relaxing the conditions on the graph topologies required in classical bearing rigidity. Under the proposed scheme in \cite{tang2021formation}, the convergence of position and velocity errors to zero can be guaranteed under much-relaxed topologies (i.e. graphs containing a single spanning tree as shown in Fig. \ref{fig:BPE}) with a single leader as opposed to the typical requirements of two leaders. However, in these works, only ideal agents with single or double-integrator models are considered. 

This work extends our previous research on formation tracking control \cite{tang2021formation} and collision avoidance in leader-follower formation control \cite{collision} to address practical challenges, such as collisions and disturbances, of controlling multiple quadrotor vehicles in realistic scenarios. We present a hierarchical control strategy for achieving formation tracking under a leader-follower directed sensing graph topology. It is assumed that only one leader quadrotor in the group knows its global position. The remaining agents, called followers, are equipped with local onboard sensors and are assumed to measure their orientation as well as the relative bearing and relative translational velocity to their neighboring agents. To ensure safe navigation in complex environments, this controller is augmented with a collision avoidance term, and a high-gain inner-loop controller is adopted for fast attitude tracking. The effectiveness of the complete system is validated through numerical simulations demonstrating the performance for time-varying shapes and rotational maneuvers, as well as experimental implementation on three physical quadrotors, considering the influence of external disturbances.

The remainder of the paper is organized as follows.
Section \ref{sec:preliminaries} provides the mathematical background on graph theory and the formal BPE definitions.
Section \ref{sec:modeling} provides the adopted quadrotor model and problem formulation, while Section \ref{sec:controlstrategy} introduces the proposed hierarchical control strategy. MATLAB simulation results are discussed in Section \ref{sec:simulations}, showing the performance for both time-varying and rigid formation shapes. Experimental results are provided in Section \ref{sec:experiments}, showing the performance and limits of the proposed method. The paper concludes with final comments in Section \ref{sec:conclusion}.

%% file: Preliminaries.tex
\section{Preliminaries} \label{sec:preliminaries}

 Let $\mathbb{S}^2 := \{y\in \mathbb{R}^3 : \lVert y \rVert = 1 \}$ denote the 2-sphere, and $\lVert . \rVert$ the Euclidean norm. The operator $\begin{bmatrix} . \end{bmatrix}_\times$ represents the skew-symmetric matrix related to any argument in $\mathbb{R}^3$. Let $I_d$ be the $d\times d$ identity matrix. We then define the projection operator $\pi_y = I_3-yy^\top \geq 0$ for any $y\in\mathbb{S}^2$, inducing $\pi_y = -[y]_\times [y]_\times$.
 
The considered formations concern systems of $n$ ($n\geq2$) connected agents. The topology between those agents can be modeled as a directed graph $\mathcal{G} := (\mathcal{V},\mathcal{E})$, where $\mathcal{V}=\{1,2,...,n\}$ is the set of vertices and $\mathcal{E} \subseteq \mathcal{V} \times \mathcal{V}$ is the set of directed edges. If the ordered pair $(i,j) \in \mathcal{E}$, then agent $i$ can sense information about the neighboring agent $j$. The set of neighbors of agent $i$ is denoted by $\mathcal{N}_i := \{j \in \mathcal{V}|(i,j)\in \mathcal{E}\}$. $N_i = |\mathcal{N}_i|$ is defined as the cardinality of the set. A digraph $\mathcal{G}$ is called an acyclic digraph if it has no directed cycle. The digraph $\mathcal{G}$ is called a directed tree with a root vertex $i, i \in \mathcal{V}$, if for any vertex $j \neq i, j \in \mathcal{V}$, there exists only one directed path connecting $j$ to $i$. 

\begin{defi} \label{defi:PEmatrix}
    A positive semi-definite matrix $\Sigma(t) \in \mathbb{R}^{n \times n}$ is called \textit{persistently exciting} (PE) if there exist $T>0$ and $0<\mu<T$ such that for all $t>0$
    \begin{equation}
        \frac{1}{T}\int_t^{t+T}\Sigma(\tau)d\tau \geq \mu I_3.
        \label{eq:PEmatrix}
    \end{equation}
\end{defi}

\begin{defi} \label{defi:PEdirection}
     A direction $y(t)\in\mathbb{S}^2$ is said to be PE if the matrix $\pi_{y(t)}$ satisfies condition \eqref{eq:PEmatrix} \cite{tang2021formation}.
\end{defi}

%% file: Modeling.tex
\section{Modeling and problem formulation} \label{sec:modeling}
\begin{figure}[t]
        \vspace{0.2cm}
	\centering
	\includegraphics[width=0.40\textwidth]{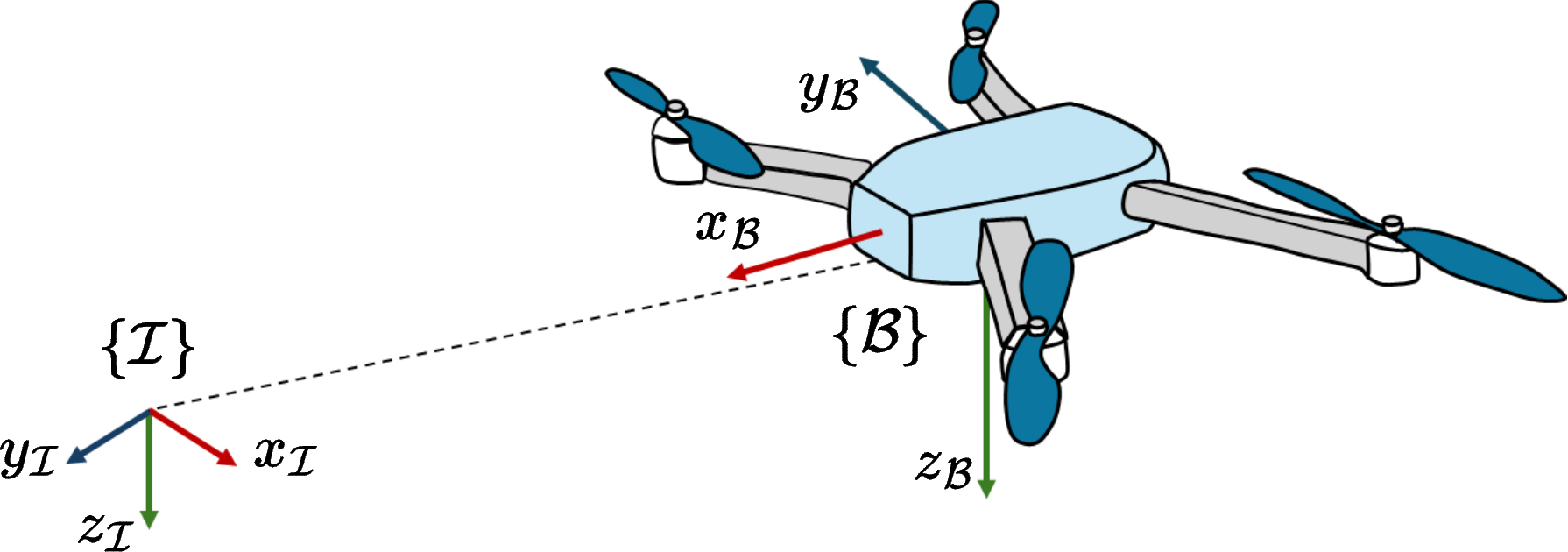}
	\caption{Schematic representation of the FRD vehicle body frame $\{B\}$, relative to an NED inertial frame $\{I\}$.}
	\label{fig:coordinates}
\end{figure}

Consider the problem of formation tracking control of a group of $n$ quadrotor vehicles. For each quadrotor $i$, let $p_i\in \mathbb R^3$ and $v_i\in \mathbb R^3$ denote its position and velocity, respectively, expressed in a common inertial frame $\{\mathcal I\}$ that follows a north-east-down (NED) convention. Let $\{\mathcal B_i\}$ be a body-fixed frame, attached to the $i$th quadrotor, following a front-right-down (FRD) convention, as shown in Fig. \ref{fig:coordinates}.
Let $R_i\in SO(3)$ denote the attitude $\{\mathcal B_i\}$ with respect to $\{\mathcal I\}$. The dynamics of quadrotor $i$ are given by
\begin{subequations}
    \label{eq:trans_dyn}
    \begin{align}
        \dot{p}_i &= v_i, \label{eq:dotp}\\
        m_i \dot{v}_i &= - T_i R_i e_3 + m_i g e_3, \label{eq:dotv} \\
         \dot{R}_i &= R_i {[\Omega_i]}_\times,\label{eq:dotR}
    \end{align}
\end{subequations}
where $g \approx 9.81\text{ms}^{-2}$ is the gravitational acceleration, $e_3 = \begin{bmatrix} 0 & 0 & 1\end{bmatrix}^\top$, $T_i\in \mathbb R_0^{+}$ is the total thrust magnitude, $m_i\in \mathbb R^{+}$ is the mass and $\Omega_i\in \mathbb R^3$ denotes the angular velocity input of agent $i$ expressed in $\{\mathcal B_i\}$.

Define the relative position vector $p_{ij}$ and relative velocity vector $v_{ij}$ between vehicle $i$ and its neighbor vehicle $j$ as 
$$p_{ij}=p_j-p_i, \quad v_{ij}=v_j-v_i, \quad \forall j\in \mathcal N_i.$$
As long as $\|p_{ij}\|\ne 0$, the relative bearing $g_{ij}$ of agent $i$ to agent $j$ is defined by
\begin{eqnarray}
    g_{ij} := \frac{p_{ij}}{\lVert p_{ij}\rVert} \in \mathbb S^2. \label{eq:bearing}
\end{eqnarray}
 Let the stacked vector $\boldsymbol{p}=[p_1^\top,...,p_n^\top]^\top\in \mathbb{R}^{3n}$ denote the configuration of $\mathcal{G}$ and the digraph $\mathcal{G}$ together with the configuration $\boldsymbol{p}$ define a formation $\mathcal{G}(\boldsymbol p)$ in three-dimensional space. We now introduce the first assumption related to the sensing graph topology.

\begin{asu} \label{asu:topology} 
    The sensing topology of the formation $\mathcal G (\boldsymbol p)$ is described as a leader-follower structure, i.e., an acyclic digraph $\mathcal{G}(\mathcal{V}, \mathcal{E})$ that has a single directed spanning tree. 
    Without loss of generality, agents are numbered (or can be renumbered) such that agent $1$ is the leader, i.e.  $\mathcal{N}_1= \varnothing$,  all other agents $i, \ i\ge 2$ are followers whose neighboring set is $\mathcal{N}_i  \subseteq \{1, \ldots, i-1\}$, according to the examples in Fig. \ref{fig:BPE}.
    Each agent $i \geq 2$ can measure the relative bearing vectors $g_{ij}$ and relative velocities $v_{ij}$ to its neighbors $j \in \mathcal{N}_i$, as well as its own attitude $R_i$. 
\end{asu}

Given this leader-follower structure, we introduce the definition of a BPE formation \cite{tang2021formation} which will be used later to define desired trajectories.
\begin{defi} \label{def:BPE}
    A leader–follower formation $\mathcal{G}(\boldsymbol p(t))$ is called BPE, if $\forall i \in \mathcal{V}$, the matrices $\sum_{j\in \mathcal{N}_i} \pi_{g_{ij}(t)}$ satisfy the PE condition \eqref{eq:PEmatrix}.
\end{defi}
Note that if $\sum_{j\in \mathcal{N}_i} \pi_{g_{ij}(t)}$ is PE, then agent $i$ has at least one bearing measurement $g_{ij}$ that is time-varying or at least two bearings that are non-collinear \cite[Lemma 1]{tang2021formation}. We now define the assumptions for the desired trajectories.
\begin{asu} \label{asu:boundedBPE}
    The desired velocity $v_{i}^*(t)$, the desired acceleration $u_i^*(t)$, and the desired jerk $\dot{u}_i^*(t)$ are bounded for all $t > 0, i \in \mathcal V$, and such that the resulting desired bearings $g_{ij}^*(t), (i,j)\in \mathcal E$ are well-defined and the desired formation is BPE for all $t > 0$.
\end{asu}

Let us now present the considered problem formulation.
\begin{prob} \label{prob:statement}
Design distributed formation tracking controllers for all follower vehicles ($i\ge 2$) such that a group of $n\ (n\ge2)$ quadrotor vehicles successfully tracks a desired BPE formation under Assumptions \ref{asu:topology}-\ref{asu:boundedBPE}, while avoiding inter-agent collisions.
\end{prob}

%% file: HierarchicalStrategy.tex
\section{Hierarchical control structure} \label{sec:controlstrategy}
\begin{figure}
	\centering
	\includegraphics[width=0.99\linewidth]{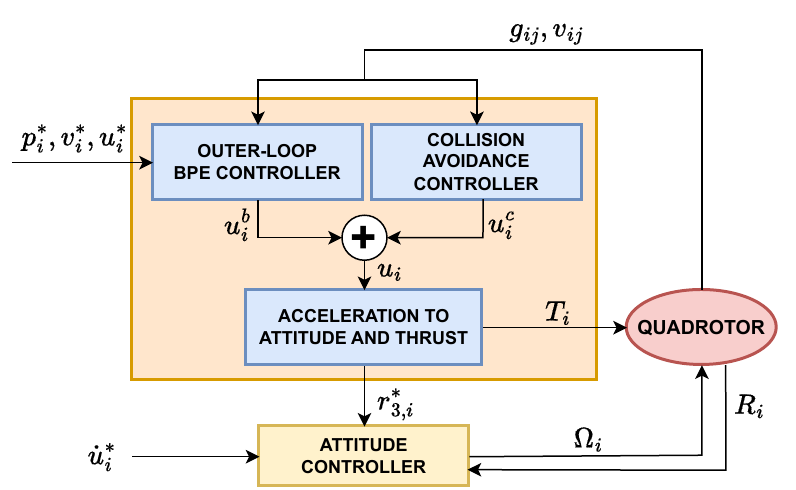}
    \vspace{-0.4cm}
	\caption{Hierarchical control structure for each quadrotor $i$.}
	\label{fig:controllayout}
\end{figure}

In this section, a hierarchical control structure (see Fig. \ref{fig:controllayout}) is proposed, assuming a time-scale separation between the translational and orientation dynamics \cite{Bertrand2011,herisse12}. Expanding the system dynamics \eqref{eq:dotv} with the desired attitude terms yields
\begin{equation}
	m_i\dot{v}_i = -T_i (R_ie_3- R_i^* e_3) - T_i R_i^* e_3 + m_i g e_3,
\end{equation}
where $-T_iR_i^* e_3$ is a total desired force. Provided that an inner-loop attitude controller is designed to provide a fast enough convergence, the time-scale separation allows for neglecting the attitude error in the outer-loop dynamics, allowing the system to be given by  
\begin{equation}
	\dot{v}_i \approx \underbrace{- \frac{T_i}{m_i} r_{3,i}^* + g e_3}_{u_i},
	\label{eqn:double_integrator_model}
\end{equation}
where $u_i \in \mathbb{R}^3$ is a virtual acceleration input to be designed and $r_{3,i}^*:=R_i^* e_3$ is the desired z-axis of the vehicle. The total thrust $T_i$ applied to each vehicle can be computed according to
\begin{equation}
	T_i := m_i||u_i - g e_3||.
\end{equation}
The desired z-axis of each vehicle is set as
\begin{equation}
	r_{3,i}^* := \frac{u_i - g e_3}{||u_i - g e_3||},
\end{equation}
from which $R_i^*\in SO(3)$ can be obtained by complementing $r_{3,i}^*$ with a desired yaw angle. Finally, the angular velocity used as input to the vehicle, adopted from \cite{Tang2015}, is given by
\begin{equation}\label{eq:Omega}
	\Omega_i := n_i [e_3]_{\times} R_i^\top r_{3,i}^* + \frac{m_i}{T_i} \pi_{e_3} R_i^\top [r_{3,i}^*]_{\times}\dot{u}_i,
\end{equation}
where $n_i \in \mathbb{R}^{+}$ is a controller gain and $ r_{3,i}$ is the third column of $R_i$. 


%% file: BPEformation.tex
\subsection{Bearing-based Formation Tracking Control} \label{sec:BPEcontrol}
Denote $\tilde{p}_{ij} = (p_{j} - p_{i}) - (p_{j}^* - p_{i}^*)$ and $\tilde{v}_{ij} = (v_{j} - v_{i}) - (v_{j}^* - v_{i}^*)$ as the respective relative position and velocity error $\forall i\ge 2$. The distributed bearing formation controller is defined as
\begin{equation}
    u_i^{b} = \sum_{j \in \mathcal{N}_i}( -{k_{p,i}} \pi_{g_{ij}} p_{ij}^*) - {k_{d,i}} \tilde{v}_{ij} + u_i^*, \ i\ge 2,
    \label{eq:BPElaw}
\end{equation}
where $u_i^{*} \in \mathbb{R}^{3}$ is a desired acceleration to be tracked, and $k_{p,i}$ and $k_{d,i}$ are positive scalar gains such that $k_{d,i} > 1$ and $k_{p,i} < \frac{4}{N_i} - \frac{4}{k_{d,i}^2 N_i^3}$.
\begin{theorem} \label{th:formation}
Consider a system with $n$ ($n\geq2$) quadrotor UAVs. For all agents $i\ge 2$, consider the closed-loop system \eqref{eqn:double_integrator_model} along with the proposed controller \eqref{eq:BPElaw}, such that $u_i = u_i^b$. If Assumptions \ref{asu:topology} and \ref{asu:boundedBPE} are satisfied, the equilibrium point $(\tilde{p}_{ij},\tilde{v}_{ij})=(0,0), \ i \ge 2$ is uniformly exponentially (UE) stable.
\end{theorem}
Considering the assumption in (\ref{eqn:double_integrator_model}), the UE stabilization of the equilibrium point $(\tilde{p}_{ij},\tilde{v}_{ij})=(0,0), \ i \ge 2$ is equivalent to the proof by mathematical induction in \cite[Theorem 2]{tang2021formation}. 

\subsection{Reactive multi-vehicle collision avoidance}
To ensure safe navigation and prevent collisions between vehicles, we propose an additive term to the bearing formation controller, leveraging the constructive barrier feedback proposed in \cite{collision}
\begin{equation}
	u_i = u_i^{b} + u_i^{c},
\end{equation}
with
\begin{equation}
	u_i^{c} = \sum_{j\in \mathcal N_i} k_{o,i}\gamma(d_{ij}) g_{ij}g_{ij}^\top v_{ij},
\end{equation}
where $d_{ij}=||p_{ij}|| - r$ and $r\in \mathbb{R}^{+}$ is the safety margin between neighboring agents and the collision avoidance effect is only active when two vehicles are close enough such that $\gamma(.)$ is chosen as
\begin{equation}
\gamma(z)=\left\{ \begin{aligned}
    0, &\ \ z\in (\epsilon_1,\infty)\\
    \phi(z) \cdot z^{-1}, & \ \ z\in [\epsilon_2,\epsilon_1]\\
    z^{-1}, &\ \ z\in (0,\epsilon_2)
\end{aligned}\right.,
\end{equation}
where $0<\epsilon_1<\epsilon_2$ and $\phi$ is a smooth function such that $\gamma$ is a continuously differentiable function for all $z\in(0,\infty)$. In this work, we use $\phi(z)=\frac{1}{2}-\frac{1}{2}\cos(\pi\frac{z-\epsilon_1}{\epsilon_2-\epsilon_1})$.
The constructive barrier feedback $u_i^c$ decreases the relative velocity in the direction of neighboring quadrotors without compromising
the performance of the nominal controller.


%% file: Simulations.tex
\section{Simulation results} \label{sec:simulations}
\begin{figure}[p]
 	\centering
 	\includegraphics[width=0.99\linewidth]{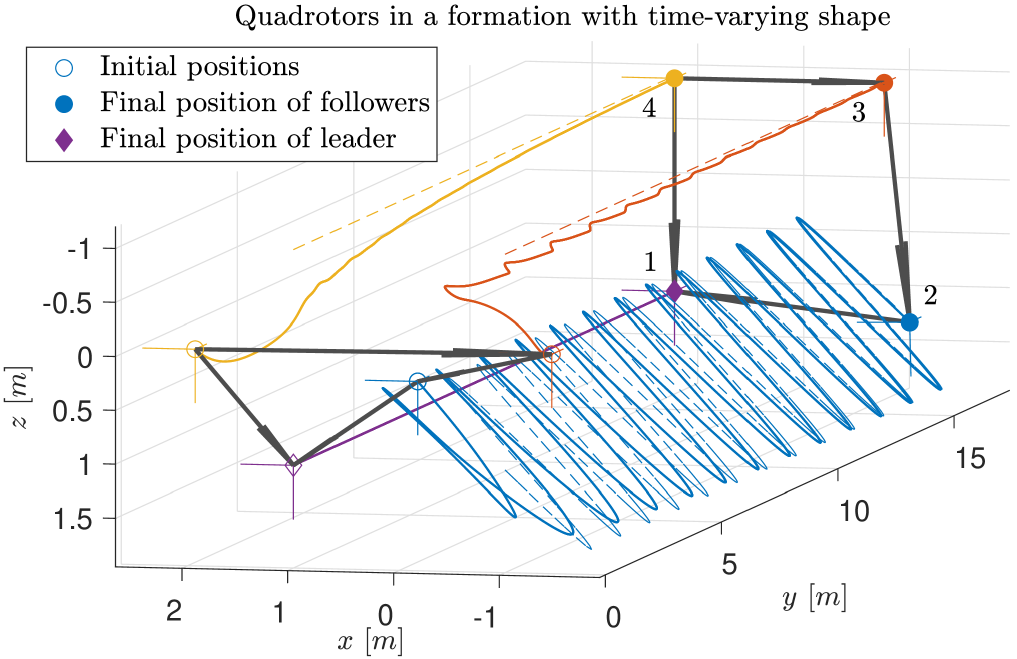}
        \vspace{-0.4cm}
 	\caption{Simulated trajectories of 4-agent formation with time-varying shape. The dashed lines represent the desired trajectories, and the solid lines the simulated trajectories. The solid black arrows indicate connections between agents.}
 	\label{fig:simvarying}
\end{figure}

\begin{figure}
	\centering
	\includegraphics[width=0.99\linewidth]{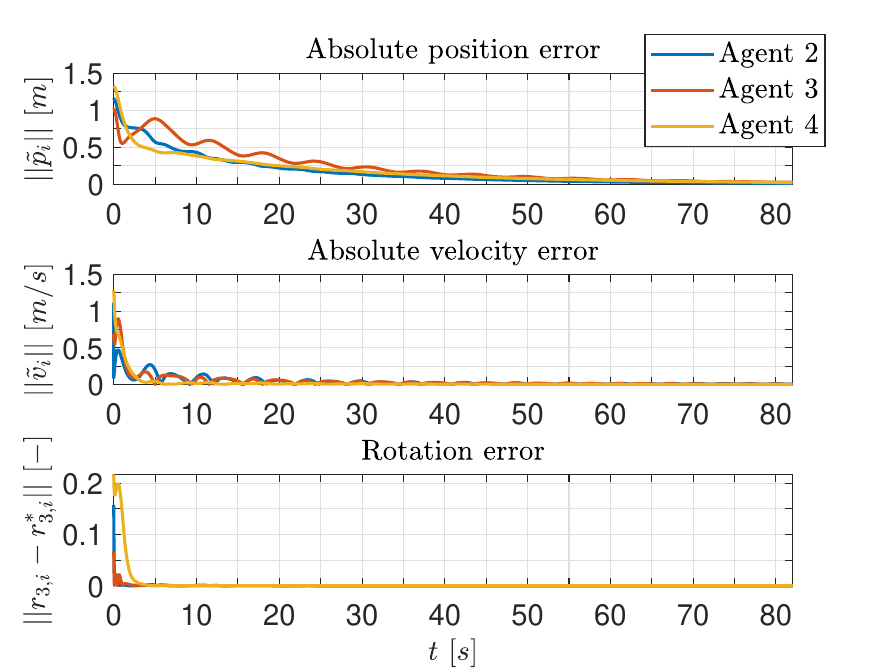}
    \vspace{-0.4cm}
	\caption{Absolute position, velocity, and rotation error of the follower agents in the 4-agent formation with time-varying shape.}
	\label{fig:simerrvarying}
\end{figure}

To validate the proposed control laws, in this section two MATLAB simulation results are provided, demonstrating the successful tracking of time-varying formations by a group of four quadrotor vehicles. The control gains adopted for both scenarios are $k_{p,i}=1.9$ and $k_{d,i}=3$, and $n_i=20, $ $\forall i\ge 2$. In the first scenario, the desired formation is chosen to have a time-varying shape while simultaneously translating along the $y$-axis. Specifically, we choose $p_1^*{=}\begin{bmatrix} 1 ~ \frac{1}{5}t ~ 1\end{bmatrix}^\top$, $p_2^*{=}\begin{bmatrix} -1{-}\frac{3}{4}\sin(t) \ \ \frac{1}{5}t \ \ 1{+}\frac{3}{4}\sin(t) \end{bmatrix}^\top$, $p_3^*{=}\begin{bmatrix} {-}1 ~ \frac{1}{5}t 
 ~ {-}1\end{bmatrix}^\top$, $p_4^*{=}\begin{bmatrix} 1 ~ \frac{1}{5}t ~ {-}1\end{bmatrix}^\top$. The three-dimensional trajectories of the formation are shown in Fig. \ref{fig:simvarying} and the evolution of state errors converging to the origin is shown in Fig. \ref{fig:simerrvarying}. The topology is described as $\mathcal{N}_2 = \{1\}$, $\mathcal{N}_3 = \{2\}$, and $\mathcal{N}_4 = \{2, 3\}$.

\begin{figure}
	\centering
	\includegraphics[width=0.99\linewidth]{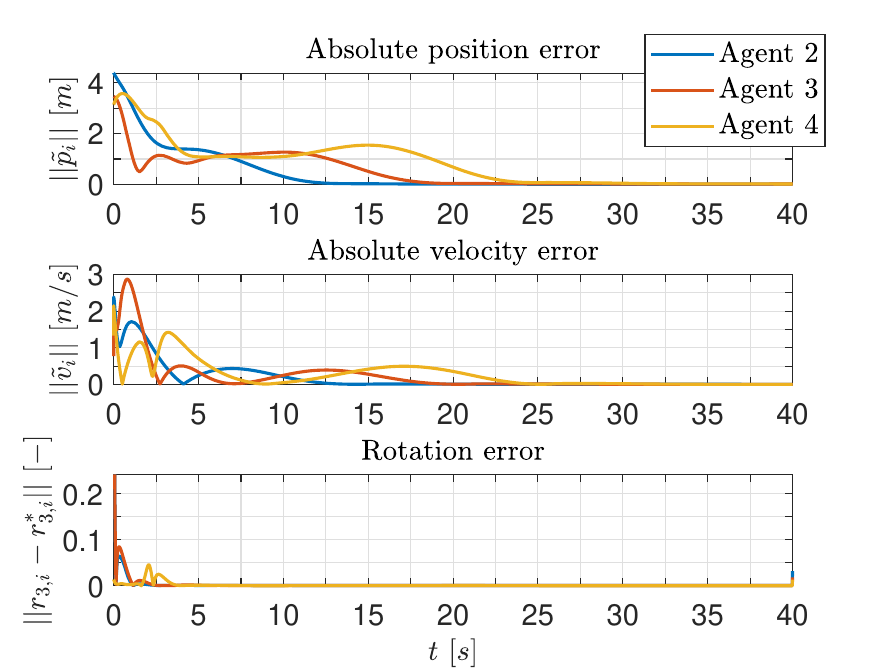}
    \vspace{-0.4cm}
	\caption{Absolute position and velocity error of the follower agents for the 4-agent formation with rigid shape, but a varying scaling factor.}
	\label{fig:simerrrescale}
\end{figure}

\begin{figure}
    \vspace{0.2cm}
    \centering
    \includegraphics[width=0.99\linewidth]{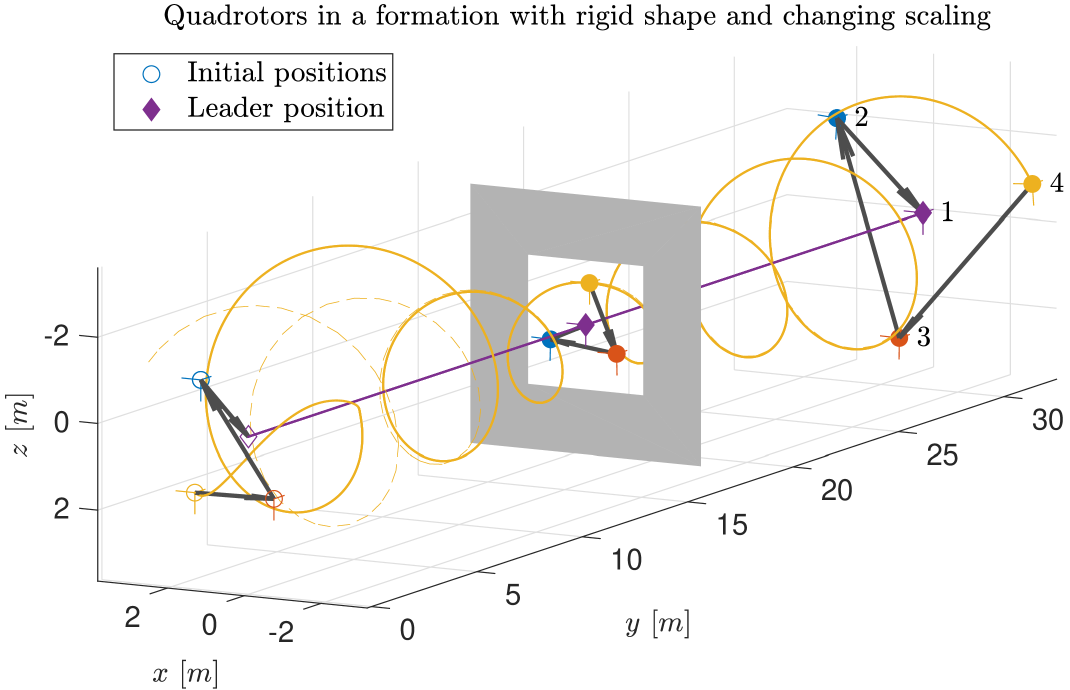}
    \vspace{-0.25cm}
    \caption{Simulated trajectories of 4-agent formation with rotating rigid shape, but a varying scaling factor. Three instances of the formation are shown, the first one (open circles) represents the initial positions, the second one at $t=40$ (solid circles) represents the point where the radius is at the smallest, after which the agents follow the desired trajectory until $t=80$ where the final position (solid circles) is shown. The dashed lines represent the desired trajectory of agent 4, and the solid yellow line is its simulated trajectory.}
    \label{fig:simrescale}
\end{figure}

In the second scenario, the desired formation is chosen to be a rigid shape rotating around and translating along the $y$-axis. At the same time, to pass through a narrow window, the formation is able to change its scale accordingly, as shown in Fig. \ref{fig:simrescale}. The formation has a minimal leader-follower graph formed by a single directed path, i.e. each follower has only one neighbor such that $\mathcal{N}_i = \{i - 1\}, i\in \mathcal{V}\setminus\{1\}$. The desired trajectories are given by $p_1^*{=}\begin{bmatrix} 0 ~ \frac{2}{5}t ~ 0\end{bmatrix}^\top$, $p_i^* = p_1^* + R_y(\frac{1}{2}t)d_i$ for $i\in\{2,3,4\}$, $d_2 = (1{+}|\frac{t-40}{20}| ) e_3$, $d_3 = R_y({-}\frac{2}{3}\pi) d_2$, $d_4 = R_y({-}\frac{4}{3}\pi) d_2$, where $R_y(\theta)$ is the rotation matrix around the $y$-axis.
The performance for this setup is shown in Fig. \ref{fig:simerrrescale}. The cascaded nature of the system is clearly visible, with the slowest convergence for agent $4$.
The numerical simulations indicate an effective performance of the proposed controllers and a clear convergence of the formation tracking errors under a variety of desired formation trajectories with more relaxed graph topologies compared to \cite{schiano2016rigidity} and  \cite{erskine2021model}. Remark that the chosen desired trajectory is one of the important factors for the convergence rate as indicated in \cite[Theorem 3]{tang2021formation}. For instance, Fig. \ref{fig:simerrrescale} shows a faster convergence than Fig. \ref{fig:simerrvarying} even though the same control gains are used.

%% file: Experiments.tex
\section{Experimental results} \label{sec:experiments}
\begin{figure}
    \centering
    \includegraphics[width=0.99\linewidth]{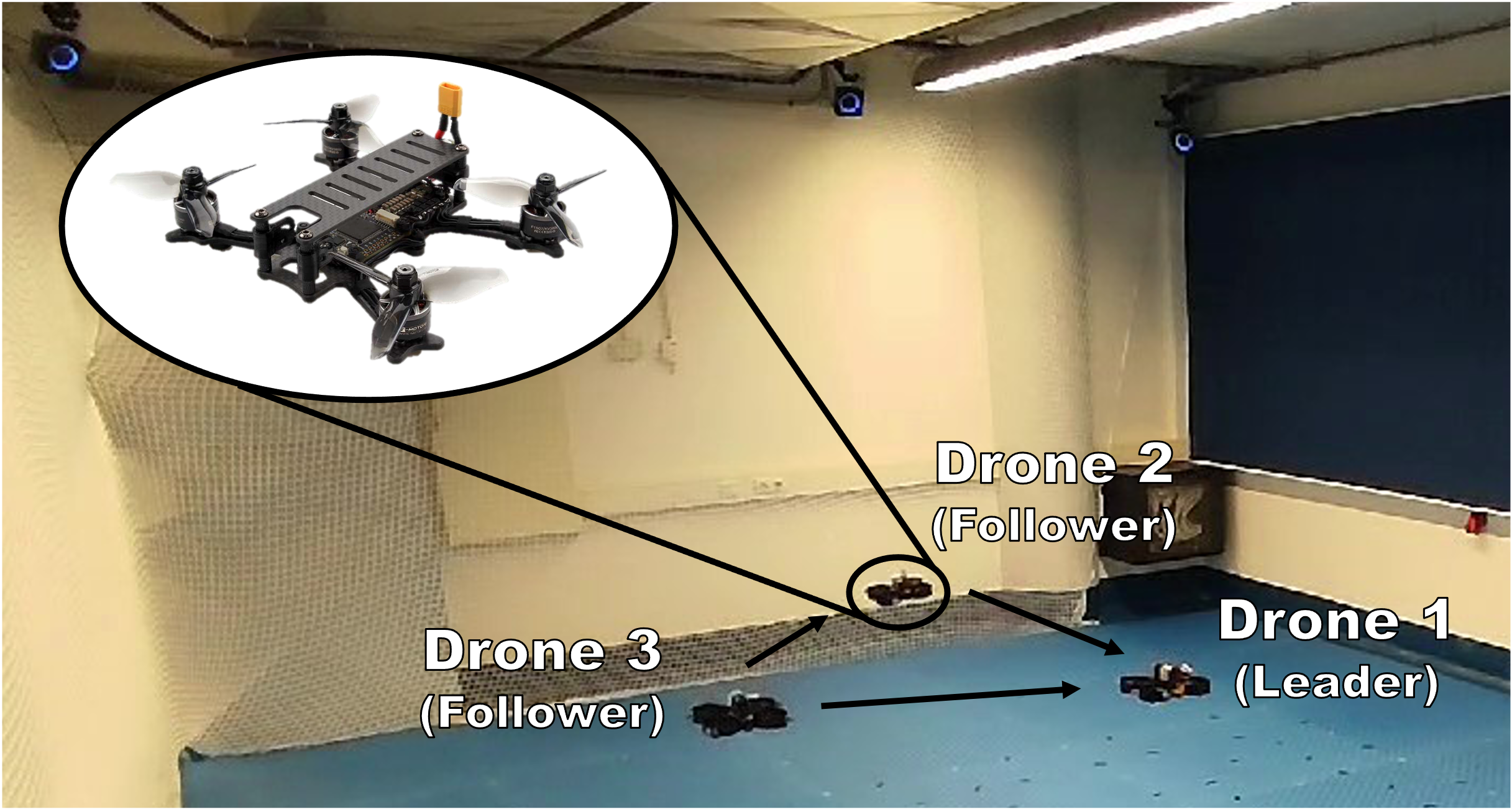}
    \vspace{-0.5cm}
    \caption{Three Kopis CineWhoop 3\textquotesingle\textquotesingle\, quadrotors during an experiment.}
    \label{fig:expshot}
\end{figure}

In this section, we describe the practical experiments carried out with three Kopis CineWhoop 3\textquotesingle\textquotesingle\, quadrotors in a closed arena (3m x 6m x 2m), with the end goal of validating the effective performance of the control laws in the presence of disturbances and communication delays, according to Fig. \ref{fig:expshot}. A setup with an OptiTrack Motion Capture (MOCAP) system is used to obtain the inter-agent virtual bearing measurements (which is a usual setup as in the previous literature \cite{schiano2016rigidity,erskine2021model}), focusing on the performance of the control laws, instead of on the acquisition of bearing measurements with a vision system. 
Each quadrotor is controlled in real time by a central desktop using UDP communication at 33 Hz over WiFi, using the Pegasus GNC with ROS 2 and PX4 as the base software infrastructure \cite{10556959}. The \href{https://github.com/SDoodeman/bpe_quadrotor}{code repository used in this experimental setup is made publicly available} and provides additional Gazebo simulations.

\begin{figure}
    \centering
    \vspace{0.2cm}
    \includegraphics[width=0.99\linewidth]{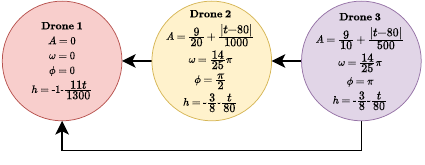}
    \vspace{-0.5cm}
    \caption{Used topology for the experiment, where, for each quadrotor, the trajectory is given by $p^*(t) = \left[
        A(t)\cos(\omega t-\phi) \ \ A(t)\sin(\omega t-\phi) \ \ h(t)
    \right]^\top$, creating a circular motion, moving upwards, and decreasing and increasing the circle radius. The limited space of the experimental arena is considered for these trajectories.}
    \label{fig:exptopology}
\end{figure}

\begin{figure}
    \centering
    \includegraphics[width=0.99\linewidth]{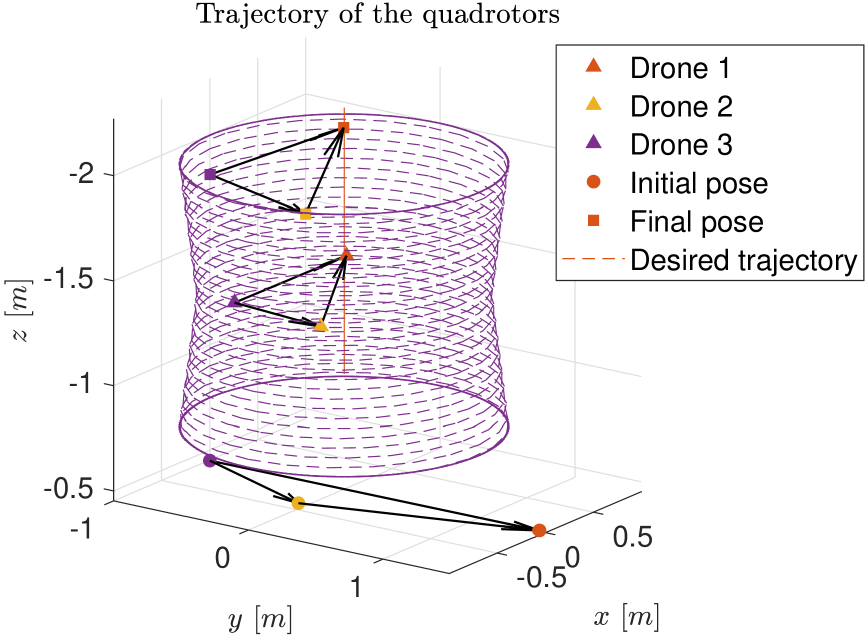}
    \vspace{-0.5cm}
    \caption{Trajectory of the Kopis quadrotors, using $k_{p} = 5.5$ and $k_{d} = 5.2$.}
    \label{fig:exp13D}
\end{figure}

\begin{figure}
	\centering
        \vspace{0.2cm}
	\includegraphics[width=0.99\linewidth]{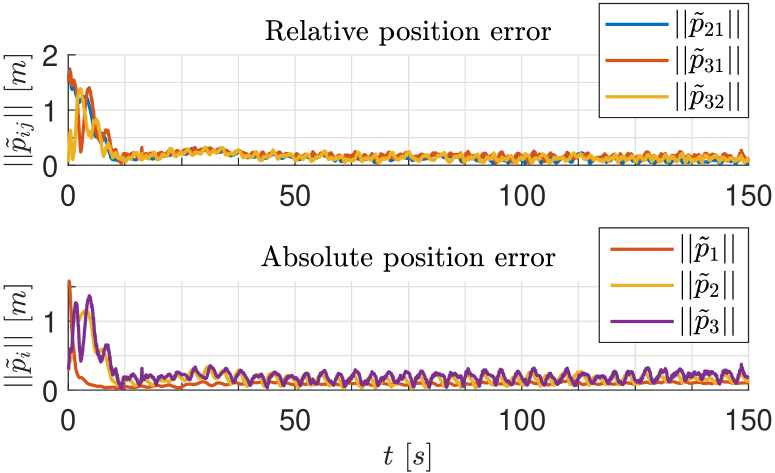}
    \vspace{-0.5cm}
	\caption{Relative position and velocity error during the experiment.}
	\label{fig:exp1err}
\end{figure}

In the practical experiments, the control gains and parameter are chosen as $k_{p,i}=5.5$ and $k_{d,i}=5.2$, $k_{o,i}=0.4$, $n_i=5.0, \forall i\ge 2$ and $r=0.10$. The adopted topology is composed of a leader quadrotor, a first follower that can only measure the relative bearing to the leader $\mathcal{N}_2 = \{1\}$, and a second follower quadrotor that is able to measure the bearing with respect to the other two vehicles $\mathcal{N}_3 = \{1, 2\}$, according to Fig. \ref{fig:exptopology}. Fig. \ref{fig:exp13D} shows the evolution of the formation in 3D space. The scale of the formation is well maintained, while only one bearing measurement to one neighbor is necessary for a following quadrotor (i.e. quadrotor 2 in Fig. \ref{fig:exptopology}), unlike the work in \cite{schiano2016rigidity} and \cite{erskine2021model} which requires one distance measurement.  Fig. \ref{fig:exp1err} shows the evolution of the state's errors, indicating the effective performance and robustness of the proposed methods on a real quadrotor formation with the presence of aerodynamic influences and disturbances.

\begin{figure}
	\centering
	\includegraphics[width=0.99\linewidth]{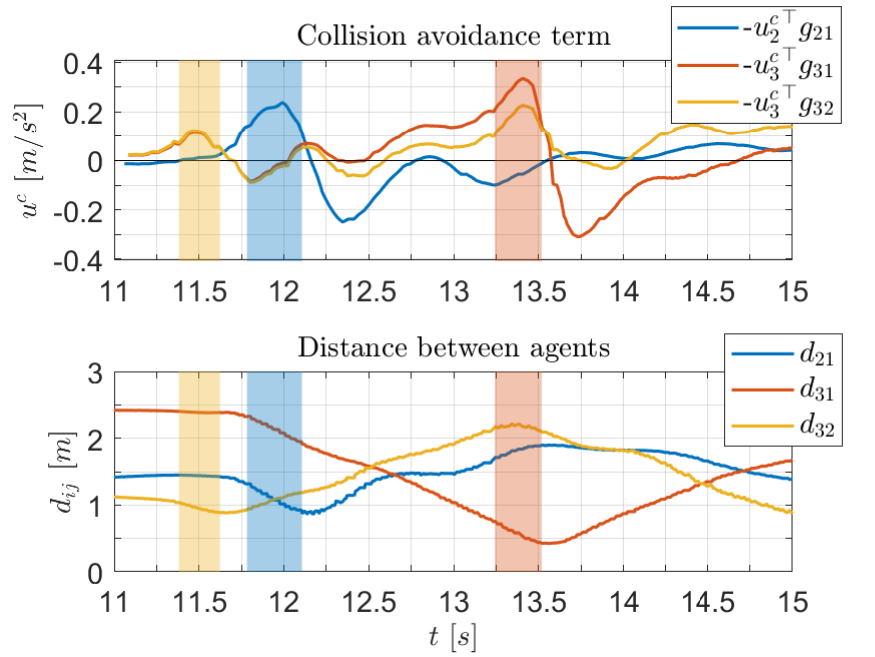}
    \vspace{-0.5cm}
    \caption{Collision avoidance term in the opposite direction to neighboring agents $-{u_i^c}^\top g_{ij}$ together with $d_{ij}$ between agents during an experiment.}
	\label{fig:coll}
\end{figure}

The effect of the collision avoidance term $u_i^c$ during the initial convergence of the vehicles to their desired formation can be seen in Fig. \ref{fig:coll}, indicating that when the distance between two agents decreases, the collision avoidance term in the opposite direction to this neighboring agent increases.

%

%% file: Conclusion.tex
\section{Conclusion} \label{sec:conclusion}
This paper presents a solution to the distributed bearing formation tracking control problem of multiple quadrotor vehicles with a limited sensor set, where only one leader knows its global position. Building upon our previous work on formation tracking control under relaxed graph topologies and leader-follower formation control with collision avoidance, we have expanded this theory to a team of quadrotors by developing a hierarchical control strategy using only inter-agent bearing measurements. This approach enables precise tracking of time-varying trajectories under relaxed graph topologies, requiring each quadrotor to connect with as few as one neighbor. The proposed method was validated by numerical simulations and practical experiments with three drones. Here, the effectiveness of the proposed method is validated even in the presence of disturbances. Future work includes implementing a controller using onboard vision systems while considering a limited field of view.